\documentclass{iau}

\usepackage{amsmath}
\usepackage{graphicx}
\usepackage{multirow}

\begin{document}

\lefttitle{Cambridge Author}
\righttitle{Proceedings of the International Astronomical Union: \LaTeX\ Guidelines for~authors}

\jnlPage{1}{7}
\jnlDoiYr{2021}
\doival{10.1017/xxxxx}

\aopheadtitle{Proceedings IAU Symposium}
\editors{C. Sterken,  J. Hearnshaw \&  D. Valls-Gabaud, eds.}

\title{Estimating Flux Densities of Diffuse Cosmological Radio Sources Exploiting Vision Transformers}

\author{N. Sanvitale$^1$, C. Gheller$^1$, F. Vazza$^2$, F. Govoni$^3$, M. Murgia$^3$, V. Vacca$^3$}
\affiliation{$^1$ Istituto di Radio Astronomia, INAF, Via P. Gobetti 101, 40129 Bologna, Italy\\
$^2$ Dipartimento di Fisica e Astronomia, Universit\`{a} di Bologna, Via P. Gobetti 92/3, 40129 Bologna, Italy\\
$^3$ Osservatorio Astronomico di Cagliari, INAF, Via della Scienza 5, 09047 Cagliari, Italy}

\begin{abstract}
We present TUNA, a Vision-Transformer based network adapted from segmentation to flux regression for faint, diffuse radio emission. Trained on LOFAR-like mock observations derived from cosmological simulations, TUNA accurately reconstructs low surface-brightness structures, with only mild smoothing and small brightness-dependent biases. Applied to LOFAR data of the A399–A401 galaxy cluster system, it recovers the ridge not identifiable in the high resolution observation and matches the low resolution tapered map. These results indicate how TUNA can deliver automated, quantitative surface brightness estimates for diffuse extragalactic sources, enabling scalable analyses for upcoming surveys.
\end{abstract}


\maketitle

\section{Introduction}

The current and upcoming generations of radio telescopes have significantly enlarged the radio observing window and enhanced their performance in terms of sensitivity and resolution, enabling the identification and characterisation of sources that have so far been barely detectable. The Warm–Hot Intergalactic Medium (e.g. \citealt{va19} and references therein)  dominates the diffuse matter of the Universe, making it a prime astrophysical target. Its high-energy tail (ultrarelativistic cosmic rays) should emit synchrotron radiation in weak intergalactic magnetic fields, yielding extremely faint signals: bright sources are expected to be $\lesssim 10^{-4}$ Jy, near current interferometer sensitivity limits \citep{va15radio}. The observational challenge is compounded by the data volume: facilities such as LOFAR \citep{van2013lofar},  MWA \citep{tingay2013murchison}, and ASKAP \citep{hotan2021australian} deliver unprecedented sensitivity, resolution, bandwidth, and field of view, and foreshadow the SKAO’s hundreds of petabytes per year\footnote{https://www.skao.int/}. Efficient numerical methods for automated source detection, segmentation, and flux density estimation are therefore essential to process this data at scale, minimize human supervision, and avoid costly reprocessing through traditional pipelines.

This work investigate the adaptation of the TUNA network \citep{2025MNRAS.541.3479S} for estimating source brightness in radio astronomy. Trained on simulated radio images, TUNA is applied to LOFAR observations to evaluate its ability to analyze faint sources at the limits of instrumental sensitivity and in complex radio environments.

\section{Methodology}

The TUNA architecture builds on TransUNet \citep{chen2021transunet}, a hybrid model that embeds a Vision Transformer encoder within a U-Net backbone. This design couples the U-Net’s precise, fine-detail characterization with transformers’ ability to model long-range context, overcoming the limited receptive field of purely convolutional approaches. Architectural details are provided in \cite{2025MNRAS.541.3479S}. For flux regression, we replaced the the original segmentation head with a linear output layer and trained using Charbonnier loss as the objective function. Training images were generated from cosmological MHD simulations \citep{2022MNRAS.509..990G}, with 150\,MHz synchrotron emission computed in post-processing from shock-accelerated electrons at various redshifts. Emission was integrated along the line of sight from $z = 0.02$ to $z = 0.15$, and full-sky models were built by stacking the resulting maps with cosmological corrections. Over 500 lightcones were produced via random rotations, yielding $2000 \times 2000$ pixel sky images with a $1.1^\circ \times 1.1^\circ$ field of view and 2\,arcsec resolution. They represent the ground truth for the training of the network. Mock LOFAR HBA observations were simulated using WSClean’s predict mode with the synthetic sky images as input \citep{offringa-wsclean-2017}. Random noise was added via LoSiTo's noise.py script\footnote{\url{https://github.com/darafferty/losito/tree/master}}, and imaging was performed with WSClean using Uniform weighting and primary beam correction, resulting in a $5.9'' \times 5.1''$ restoring beam. These images, together with the original sky maps, constitute the training and test datasets for TUNA.

\section{Results and Discussion}

TUNA was trained on one hundred $2000 \times 2000$ pixel images, using 10\% of the data for validation. Network performance was evaluated on an independent test set of 50 additional images. Hyperparameter tuning followed the procedure described in \cite{2025MNRAS.541.3479S}, leading to the optimal configuration: learning rate $\mu = 0.0005$, batch size $B = 24$, training steps per epoch $N_T = 512$, and number of epochs $E_p = 300$.

\begin{figure}
\centering
    \includegraphics[scale=.3]{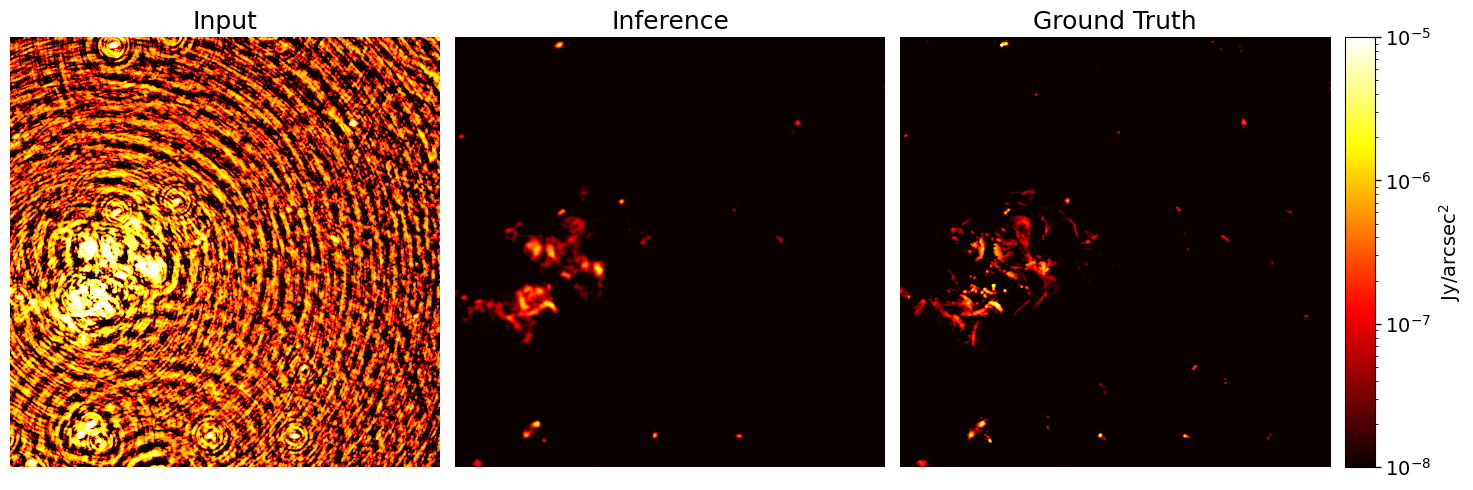}
    \caption{Example of a $1.1^\circ \times 1.1^\circ$ mock image used as input for the network (left) and the corresponding prediction from TUNA (center). The sky map (our ground truth) is shown on the right.}
    \label{fig:mock}
\end{figure}

\begin{figure}
\centering
    \includegraphics[scale=.45]{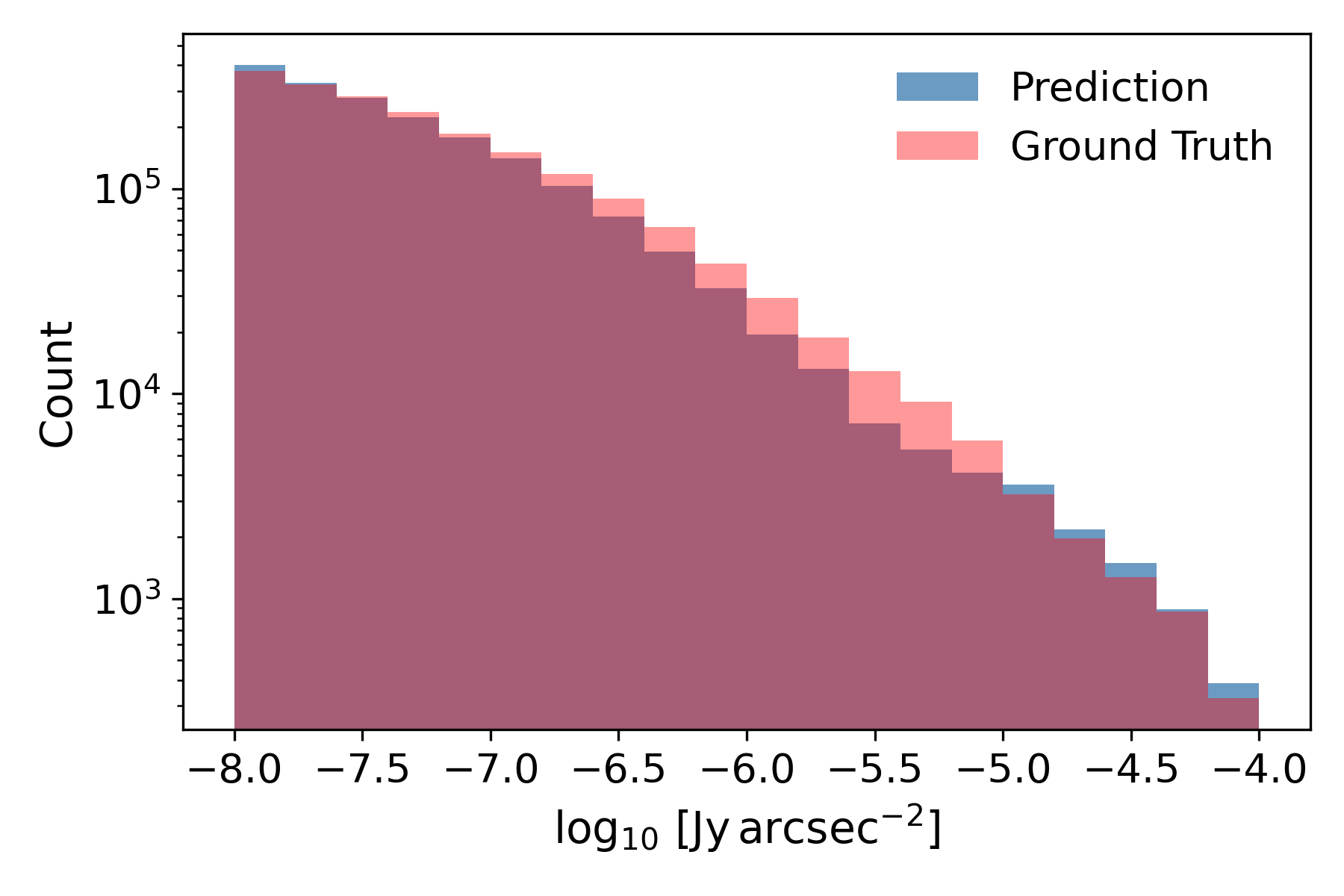}
    \caption{Distribution of surface brightness from the test and prediction data with surface brightness higher than $10^{-8}$Jy/arcsec$^2$.}
    \label{fig:histo}
\end{figure}

Figure~\ref{fig:mock} shows an example of an input image alongside with the corresponding ground-truth map utilised for training, in addition to TUNA’s predictions. The network effectively recovers diffuse low surface brightness emission despite significant imaging artifacts and noise in the input data, accurately reproducing source morphology. TUNA’s predictions exhibit mild diffusion, producing slightly blurred surface brightness maps compared to the ground truth. The network also tends to slightly overestimate emission at higher intensities. As shown in Fig.\ref{fig:histo}, the surface brightness distributions of pixels above $10^{-8}$ Jy/pixel in both the ground truth and predicted maps are broadly consistent. However, TUNA underestimates brightness in the $10^{-7}$ – $10^{-5}$ Jy/arcsec² range and overestimates it at higher emission levels.

\begin{figure}
\centering
    \includegraphics[scale=.3]{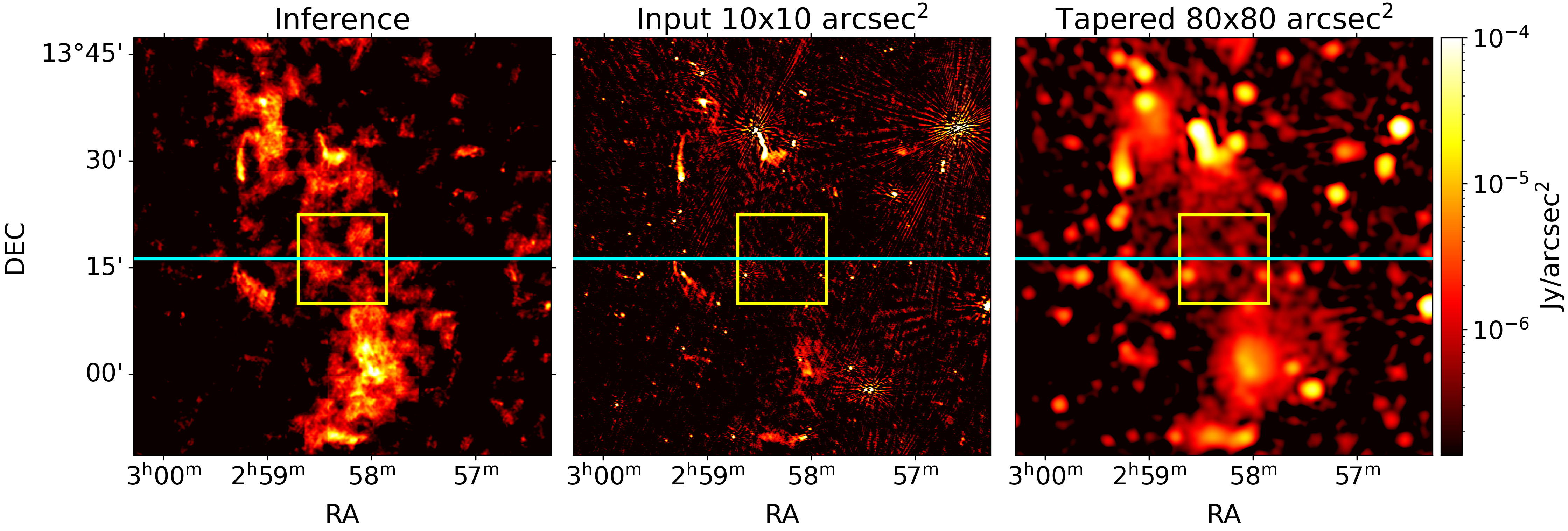}
    \caption{Predicted surface brightness distribution for the A399-A401 system (left panel) compared with the input image (centre) and the $80$ arcsec tapered image (right). All panels are thresholded at 1$\sigma$, where $\sigma$ is the rms noise measured in each image ($\sigma_\mathrm{Inference} = 1.22\times 10^{-7}$ Jy/arcsec$^2$, $\sigma_\mathrm{Input} = 2.65\times 10^{-6}$ Jy/arcsec$^2$, $\sigma_\mathrm{Tapered} = 1.38\times 10^{-7}$ Jy/arcsec$^2$). The box and cyan line indicate the regions where the flux density and the surface brightness have been evaluated.}
    \label{fig:ridges}
\end{figure}

To assess TUNA’s generalization to real observational data, the model was applied to LOFAR high band antenna (HBA) observations of the radio ridge connecting the pre-merging cluster pair Abell 399 and Abell 401 (A399-A401, \citealt{govoni2019radio}). The input was a LOFAR HBA image at a central frequency of 140 MHz, with $10$ arcsec resolution and an rms sensitivity of $\sigma = 0.3$ mJy/beam. Fig. \ref{fig:ridges}, presents the TUNA-predicted surface brightness alongside the original input image and a $80$ arcsec tapered image with sensitivity $\sigma= 1\; {\rm mJy/beam}$. The network shows the capability to effectively probe low surface brightness diffuse emission underscoring its crucial role in detecting faint and elusive sources. 

To derive an approximate flux scale for the network predictions, we used two bright, extended sources present in the observed field as empirical calibrators and applied the average of the resulting scaling factors. Systematic uncertainty was estimated from the relative difference in diffuse flux densities obtained using each calibration. The total uncertainty was then computed by combining this with the statistical error from image noise. While suitable for this preliminary analysis, a more robust and general calibration pipeline is currently under development.

\begin{figure}
\centering
    \includegraphics[scale=.37]{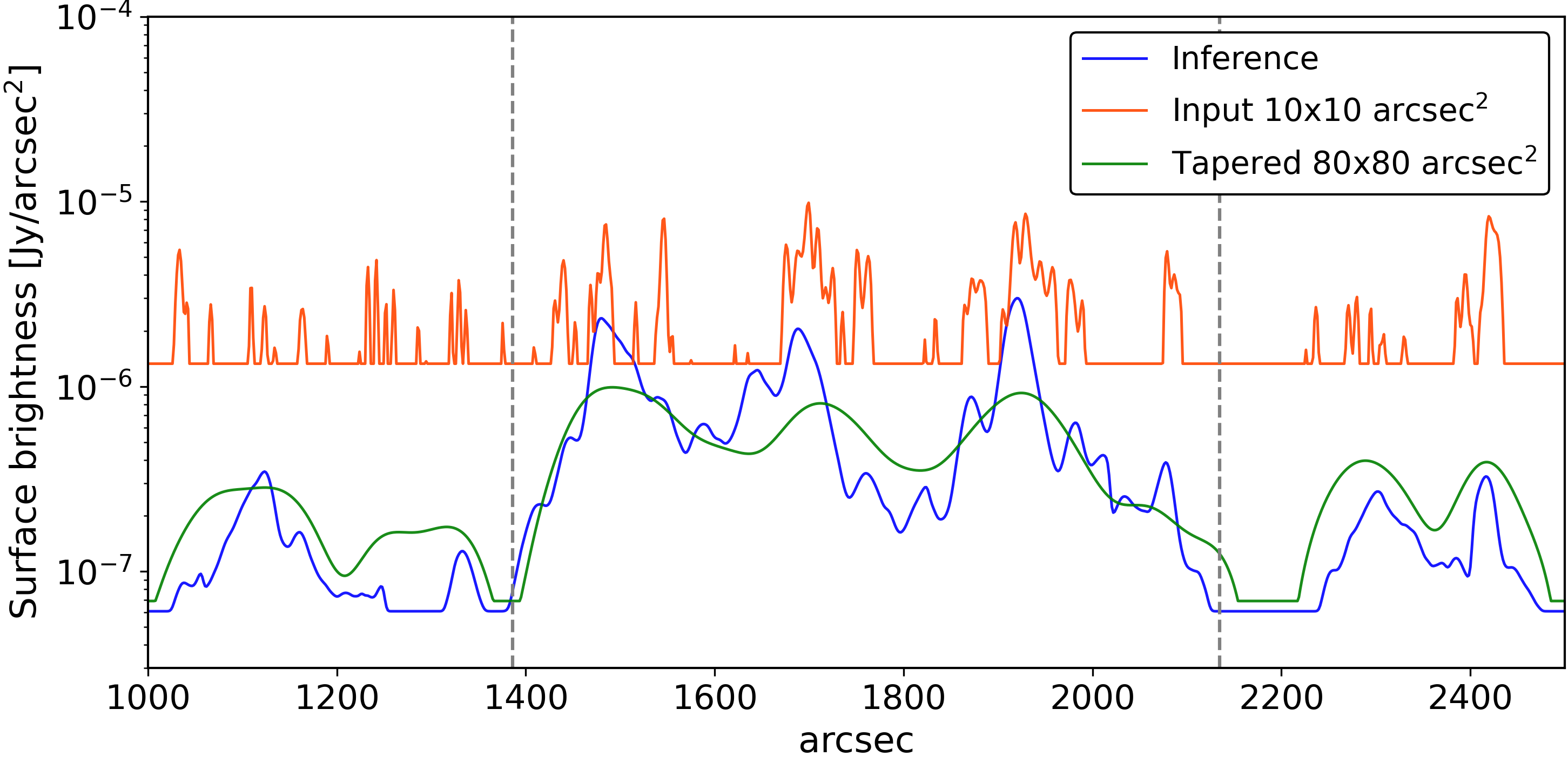}
    \caption{Zoom-in on the 1D surface brightness profile across the ridge region. The plot shows the predicted (blue), input (red), and tapered (green) surface brightness curves, each thresholded at 0.5$\sigma$. Vertical dashed lines indicate the boundaries of the yellow box used for flux integration.}
    \label{fig:cut}
\end{figure}

A cyan line marks the 1D cut used to extract a transverse surface brightness profile across the ridge (see Fig.~\ref{fig:cut}). The comparison illustrates TUNA’s ability to reconstruct diffuse emission consistent with the low-resolution tapered image, while also preserving small-scale substructures seen in the high-resolution image, despite the ridge being undetectable in that input. The yellow box in Fig. \ref{fig:ridges} outlines the region used to calculate the total flux density, resulting in $0.323\pm 0.064$ Jy for the prediction and $0.341\pm 0.051$ Jy for the for the tapered image, assuming a $15\%$ systematic uncertainty in LOFAR's flux scale calibration \cite{govoni2019radio}. The high-resolution input image, dominated by the noise, gives an upper limit for the flux density of $2.20\pm 0.193$ Jy. 

Overall, TUNA produces a surface brightness distribution that reveals the diffuse ridge otherwise undetectable in the high-resolution image. Although the result relies on a field-specific calibration and an initial implementation of the network, both aspects are being refined to improve generalization and robustness across LOFAR datasets, providing a strong basis for further development.

\section*{Acknowledgements}

This paper is supported by the Fondazione ICSC, Spoke 3 Astrophysics and Cosmos Observations. National Recovery and Resilience Plan (PNRR), Project ID CN\_00000013 "Italian Research Center for High-Performance Computing, Big Data and Quantum Computing" funded by MUR Missione 4 Componente 2 Investimento 1.4. We acknowledge the CINECA award under the ISCRA initiative (project INA24\_C5B10). 

\bibliographystyle{plainnat}
\bibliography{bibliography}

\end{document}